# Ultrasensitive Higher-Order Exceptional Points via Non-Hermitian Zero-Index Materials

Dongyang Yan[1], Alexander S. Shalin[2,4], Yongxing Wang[3,*], Yun Lai[5], Yadong Xu[1],

Zhi Hong Hang[1], Fang Cao[2], Lei Gao[1,2,*], Jie Luo[1,*]

[1]School of Physical Science and Technology & Collaborative Innovation Center of Suzhou Nano Science and Technology & Jiangsu Key Laboratory of Frontier Material Physics and Devices, Soochow University, Suzhou 215006, China

[2]School of Optical and Electronic Information, Suzhou City University & Jiangsu (Suzhou) Key Laboratory of Biophotonics, Suzhou 215104, China

[3]Zhangjiagang Campus, Jiangsu University of Science and Technology, Zhangjiagang 215600, China

[4]Center for Photonics and 2D Materials, Moscow Institute of Physics and Technology, Dolgoprudny 141700, Russia

[5]National Laboratory of Solid State Microstructures, School of Physics, Collaborative Innovation Center of Advanced Microstructures and Jiangsu Physical Science Research Center, Nanjing University, Nanjing 210093, China

*Correspondence: 201900000107@just.edu.cn (Yongxing Wang); leigao@suda.edu.cn (Lei Gao); luojie@suda.edu.cn (Jie Luo)

**Abstract:** Higher-order exceptional points (EPs) in optical structures enable ultra-sensitive responses to perturbations. However, previous investigations on higher-order EPs have predominantly focused on coupled systems, leaving their fundamental physics in open scattering systems largely unexplored. Here, we harness wave interference to realize higher-order EPs in non-Hermitian zero-index materials connected to multiple open channels. Specifically, we show that a three-channel model can give rise to three interesting types of third-order EPs: lasing EP, reflecting EP, and absorbing EP. Notably, near the third-order absorbing EP, we show ultrasensitivity − a drastic change in output power in response to perturbations at the operating frequency − in a purely lossy system. These findings pave the way for achieving higher-order and even arbitrary-order EPs in open scattering systems, offering significant potential for advanced sensing applications.

Non-Hermitian optical systems, which exploit the interplay of gain and/or loss, provide a powerful platform for exploring novel and intriguing phenomena that are absent in Hermitian systems [1-10]. A prominent property of non-Hermitian optical systems is the presence of singularities known as exceptional points (EPs), where eigenvalues and their corresponding eigenvectors coalesce [3-10]. These EPs give rise to remarkable phenomena, such as anomalous transmission or reflection [3], with promising applications in lasing [6, 11-13], absorption [14, 15], cloaking [16-18], sensing [7, 8, 19-28], etc.

In particular, higher-order EPs [29-42], arising from the coalescence of multiple eigenvalues, offer a promising avenue for significantly enhancing the sensitivity of optical systems to perturbations. An $N$th-order EP exhibits an $N$th root dependence of eigenvalue splitting on perturbations [7, 8], making them especially appealing for enhancing sensitivity in sensing and detection applications. To date, higher-order EPs and their sensing applications have been demonstrated in a variety of systems, including coupled microcavities [31-33], microdisk cavities with coupling modes [34], coupled resonant circuits or coils [35-38], loss/gain-modulated periodic coupled resonators [39-41], photonic time-Floquet crystals [42], etc. Intriguingly, even lossless periodically coupled waveguides can support higher-order EPs through Floquet-Bloch eigenwave coalescence [43-45]. Notably, EPs of arbitrary order can, in principle, be achieved by increasing the number of coupled resonators with fine-tuned parameters [32, 38-40]. However, these studies have predominantly focused on coupled systems described by Hamiltonians, where perturbation-induced frequency splitting is the primary focus. Quite interestingly, similar to Hamiltonian systems, well-defined EPs can also emerge in open scattering systems characterized by scattering matrices when two or more eigenvalues and their associated eigenvectors coalesce [4, 46]. Recent investigations have uncovered a remarkable variety of EPs, including lasing EPs [11-13], absorbing EPs [14, 15], as well as the dynamic creation and annihilation of EPs [47]. Despite these achievements, the fundamental physics and characterization of higher-order EPs based on scattering matrices, as well as their sensing applications in open scattering systems, remain rarely explored. This gap stems from the inherent challenge in open systems, which typically allow interference between two counterpropagating waves [23] but face significant hurdles in enabling interference among multiple waves incident from different directions.

In this work, we show that zero-index materials (ZIMs) [48-62], as a unique class of materials with near-zero permittivity and/or permeability, provide a solution to enable coherent interference among multiple waves incident from different directions [53-58, 62]. This establishes ZIMs as a unique platform for realizing higher-order EPs in open scattering systems. Traditionally, the material loss/gain is considered detrimental in ZIM systems, interestingly, here, we are making use of the interplay of loss and/or gain to realize higher-order EPs and sensing applications.

We show that non-Hermitian ZIMs integrated with multiple channels enable, in principle, the realization of EPs of arbitrary order in open scattering systems, depending on the number of channels. Using scattering matrices, we reveal that a three-channel model can exhibit three solutions of third-order EPs, and each solution can give rise to three interesting types of third-order EPs: lasing EP, reflecting EP, and absorbing EP. Notably, near the third-order absorbing EP, a gas sensor using purely lossy ZIMs has been numerically verified, showing a pronounced sensitivity to perturbations in refractive index of input gases at the operating frequency, highlighting its potential as an advanced sensing platform.

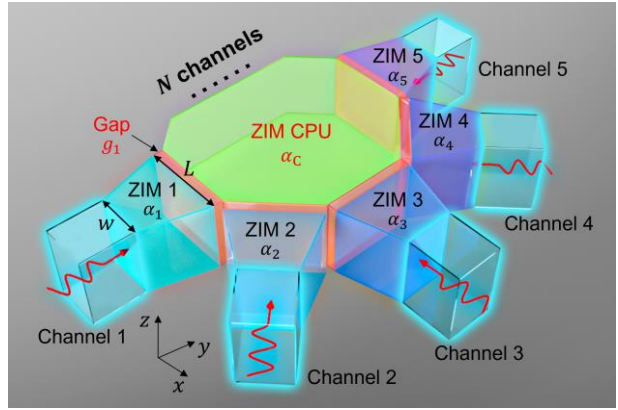

**FIG. 1.** An open non-Hermitian ZIM model for higher-order EPs. It consists of a ZIM CPU connected to $N$ channels, spaced by air gaps. Each channel comprises two parts: a hollow straight region for wave input and output, and a trapezoidal ZIM-filled region connecting the ZIM CPU. All ZIM components possess the same near-zero permittivity, but distinct complex permeabilities.

The non-Hermitian ZIM model for higher-order EPs is schematically shown in Fig. 1. It consists of a central ZIM "core", referred to as the ZIM "central processing unit (CPU)", which plays a crucial role in enabling the interference of multiple waves arriving from distinct channels. For simplicity, we model the ZIM CPU as a regular polygon with side length $L$. It is connected to $N$ channels, with an air-gap $g_n$ separating the $n$th channel from the ZIM CPU. Each channel, having a width $w$, comprises two parts: a hollow straight region for wave input and output, and a trapezoidal ZIM-filled region connecting the ZIM CPU. The ZIM CPU has an area $A_C$, and all filling ZIMs in the channels have the same area $A$. All ZIM components are characterized by near-zero permittivity (i.e., $\varepsilon_{ZIM} \approx 0$) but exhibit distinct complex permeabilities. Let $\mu_C$ (or $\mu_n$) denote the relative permeability of the ZIM CPU (or the filling ZIM in the $n$th channel).

We consider the illumination of transverse-magnetic (TM) polarization, with magnetic field oriented along the $z$ direction. Here, a time variation term of $e^{-i\omega t}$ is assumed, where $\omega$ is the angular frequency. The model's upper and lower boundaries are set as perfect magnetic conductors, while its side boundaries are perfect electric conductors, enabling the system to be reduced to a two-dimensional model on the $xy$ plane. Within each channel, only the fundamental transverse electromagnetic mode characterized by magnetic field $H_z$ is allowed. The wave behaviors within this $N$-channel non-Hermitian ZIM model can be described by a scattering matrix as follows,

$$S = \begin{pmatrix} S_{11} & \cdots & S_{1N} \\ \vdots & \ddots & \vdots \\ S_{N1} & \cdots & S_{NN} \end{pmatrix}, \qquad (1)$$

where $S_{mn}$ $(m, n = 1, 2, \ldots, N)$ is the S-parameter, describing the complex-valued reflection ($S_{mm}$) or transmission ($S_{mn}$, $m \neq n$) coefficient at the waveguide ports [63]. Here, we obtain the S-parameters by calculating the ratio of output magnetic field to input magnetic field.

We emphasize that the introduction of air gaps between the ZIM CPU and each channel is critical for realizing EPs. In the absence of these air gaps, the magnetic field across all ZIM regions becomes uniform. This phenomenon can be explained based on the Ampère-Maxwell equation $\nabla \times \mathbf{H} = -i\omega\varepsilon_{ZIM}\mathbf{E}$. When $\varepsilon_{ZIM} \approx 0$, the magnetic field $H_z$ satisfies $\partial H_z/\partial x \approx \partial H_z/\partial y \approx 0$, which implies a uniform magnetic-field distribution within the ZIM regions. Considering the boundary continuity conditions of magnetic fields within each channel, we obtain the relation $1 + S_{nn} = S_{mn}$ $(m \neq n)$. This relation imposes constraints on the correlations among the S-parameters. Under this circumstance, we find that the eigenvalues of the scattering matrix are $s_1 = s_2 = \cdots = s_{N-1} = -1$ and $s_N = N - 1 + \sum_{n=1}^{N} S_{nn}$. Although $N - 1$ eigenvalues are degenerate, their eigenvectors remain orthogonal, as in Hermitian systems, precluding the emergence of EPs. Interestingly, here we disrupt the uniformity of the magnetic field across all ZIM regions through introducing air gaps, though the magnetic field within each region remains constant. This disturbance breaks the condition $1 + S_{nn} = S_{mn}$. Consequently, the air gaps offer us additional degrees of freedom to tune the S-parameters, thus facilitating the realization of higher-order EPs.

Based on the uniform magnetic field assumption within each ZIM region, boundary continuity conditions, and the Faraday-Maxwell equation, the S-parameter $S_{mn}$ can be expressed as (see Supplemental Material [64]),

$$S_{mn} = \frac{-2L/w}{[(1-i\alpha_m)\sin(k_0 g_m)+i(L/w)\cos(k_0 g_m)][(1-i\alpha_n)\sin(k_0 g_n)+i(L/w)\cos(k_0 g_n)]\gamma}$$
$$+ \delta_{mn} \frac{(1+i\alpha_m)\sin(k_0 g_m)-i(L/w)\cos(k_0 g_m)}{(1-i\alpha_m)\sin(k_0 g_m)+i(L/w)\cos(k_0 g_m)}, \tag{2}$$

where $\alpha_{m(n)} = \frac{k_0 A \mu_{m(n)}}{w}$, $\gamma = \sum_{n=1}^{N} \frac{(1-i\alpha_n)\cos(k_0 g_n)-(L/w)i\sin(k_0 g_n)}{(L/w)\cos(k_0 g_n)-i(1-i\alpha_n)\sin(k_0 g_n)} - i\alpha_C w/L$ with $\alpha_C = \frac{k_0 A_C \mu_C}{w}$. $k_0$ is wave number in free space. The quantities $\alpha_{m(n)}$ and $\alpha_C$ are characteristic parameters of ZIMs, encapsulating both geometric properties and electromagnetic parameters. The sign of their imaginary parts directly indicates the lossy or gain characteristic: a positive imaginary part implies loss, and a negative one implies gain. Using Eqs. (1) and (2), the scattering matrix can be explicitly determined, enabling detailed analysis of its eigenvalues and associated EPs. However, the analysis of the general case remains complex. Therefore, for simplicity, as for demonstration purpose, we focus on a three-channel model to elucidate the realization of higher-order EPs, their underlying physics, and potential sensing applications.

Figure 2(a) illustrates the top-down view of the three-channel model characterized by $\mu_C \approx 0$, $g_1 = g_2 = \lambda_0/4$ and $g_3 = 0$ ($\lambda_0$ is the free-space wavelength). In this model, three distinct solutions for third-order EPs are obtained (see Supplemental Material [64]). The degenerate eigenvalues associated with first solution are,

$$s_{1,2,3} = -\frac{\alpha_3 - (1-\sqrt{3}L/w)i}{\alpha_3 + (1+\sqrt{3}L/w)i}, \tag{3}$$

under the conditions of $\alpha_1 = \alpha_3 + (1 + 3i\sqrt{3}) L/(2w)$ and $\alpha_2 = \alpha_3 - (1 - 3i\sqrt{3}) L/(2w)$. For the second solution, the degenerate eigenvalues are,

$$s_{1,2,3} = -\frac{\alpha_3 - (1+\sqrt{3}L/w)i}{\alpha_3 + (1-\sqrt{3}L/w)i}, \tag{4}$$

under the conditions of $\alpha_1 = \alpha_3 + \left(1 - 3i\sqrt{3}\right)L/(2w)$ and $\alpha_2 = \alpha_3 - \left(1 + 3i\sqrt{3}\right)L/(2w)$. For the third solution, the degenerate eigenvalues are,

$$\text{and } s_{1,2,3} = -\frac{\alpha_3 - i}{\alpha_3 + i}. \tag{5}$$

under the conditions of $\alpha_1 = \alpha_3 + i\sqrt{2}\,L/w$ and $\alpha_2 = \alpha_3 - i\sqrt{2}\,L/w$.

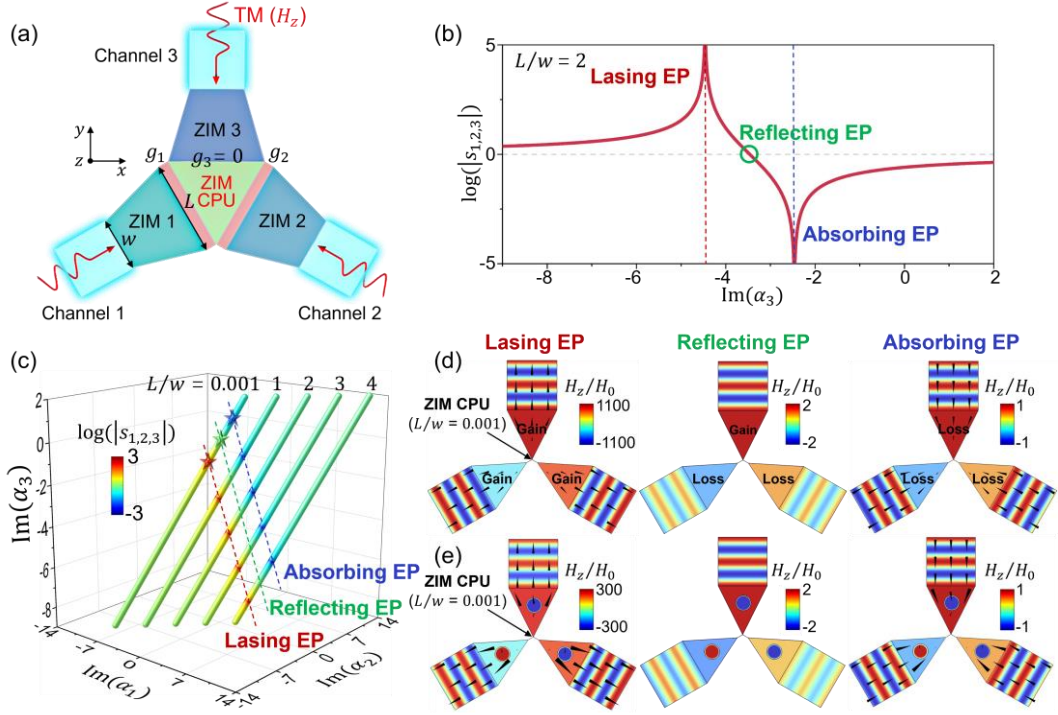

**FIG. 2.** (a) Top-down view of a three-channel non-Hermitian ZIM model. (b) Degenerate eigenvalues, $\log|s_{1,2,3}|$, for the first solution of the third-order EP [Eq. (3)] as a function of $\text{Im}(\alpha_3)$, fixing $L/w = 2$ and assuming $\alpha_3$ to be purely imaginary. (c) Evolution of the third-order EP of the first solution in the parameter space $\{\text{Im}(\alpha_1), \text{Im}(\alpha_2), \text{Im}(\alpha_3)\}$ as $L/w$ increases, where $\text{Re}(\alpha_1) = L/(2w)$, $\text{Re}(\alpha_2) = -L/(2w)$, and $\text{Re}(\alpha_3) = 0$. The color scale represents $\log|s_{1,2,3}|$. [(d) and (e)] Simulated distributions of normalized magnetic fields $H_z/H_0$ (color map) and time-averaged Poynting vectors (arrows) for the third-order lasing EP (left), reflecting EP (middle), and absorbing EP (right) when $L/w = 10^{-3}$, as marked by stars in (c). In (d), ideal non-Hermitian ZIM models are considered, while in (e), the non-Hermitian ZIMs are implemented using ENZ materials doped with circular lossy/gain dopants.

We first analyze the first solution of the third-order EPs, as expressed in Eq. (3). This solution exhibits unique eigenvalue characteristics — specifically, infinite, unity, and zero eigenvalues − corresponding to third-order lasing, reflecting, and absorbing EPs, respectively. The three unique types of third-order EPs occur when the parameter $\alpha_3$ respectively satisfies,

$$\alpha_3 = -\left(1 + \sqrt{3}\,L/w\right)i, \ \alpha_3 = -\sqrt{3}i\,L/w, \text{ and } \alpha_3 = \left(1 - \sqrt{3}\,L/w\right)i. \quad (6)$$

While lasing and absorbing phenomena at EPs have been previously reported in second-order EP systems [11-15], our findings first show that these behaviors also occur in higher-order EP systems. More interestingly, we find that the higher-order absorbing EP can give rise to enhance sensitivity, as will elucidate in the following. Additionally, we uncover a novel type of EP: the higher-order reflecting EP occurring when $s_{1,2,3} = 1$, where the outgoing power within each channel equals the input power. Under the condition of this reflecting EP, the S-parameters obey the relation $S_{m1}\psi_1 + S_{m2}\psi_2 + S_{m3}\psi_3 = \psi_m$, where $m = 1,2,3$ and $\psi_m$ is the $m$-th element of the coalesced eigenvector (see Supplemental Material [64]).

To explore the underlying physics and unique optical phenomena associated with these higher-order EPs, we plot the degenerate eigenvalues $\log|s_{1,2,3}|$ for the first solution of the third-order EPs [Eq. (3)] as the function of the imaginary part of $\alpha_3$ (i.e., $\text{Im}(\alpha_3)$), fixing $L/w = 2$ and assuming $\alpha_3$ to be purely imaginary, as shown in Fig. 2(b). The relevant parameters are $L = 4\lambda_0$, $A_C = \sqrt{3}L^2/4$, $A = 6\lambda_0^2$. Degenerate eigenvalues of infinite, unity, and zero are observed at $\alpha_3 = -\left(1 + 2\sqrt{3}\right)i$, $\alpha_3 = -2\sqrt{3}i$, and $\alpha_3 = \left(1 - 2\sqrt{3}\right)i$, respectively. These results are consistent with the theoretical predictions in Eq. (6), confirming the presence of third-order lasing, reflecting, and absorbing EPs. Figure 2(c) shows the evolution of the third-order EPs in the parameter space $\{\text{Im}(\alpha_1), \text{Im}(\alpha_2), \text{Im}(\alpha_3)\}$ as $L/w$ increases. Here, the real parts of $\alpha_1$, $\alpha_2$, and $\alpha_3$ are $\text{Re}(\alpha_1) = L/(2w)$, $\text{Re}(\alpha_2) = -L/(2w)$, and $\text{Re}(\alpha_3) = 0$, respectively, for satisfying the conditions for the first solution of the third-order EPs. The color scale represents $\log|s_{1,2,3}|$, demonstrating the existence of the three types of third-order EPs for different $L/w$.

An intriguing case with $L/w = 10^{-3}$ is further analyzed, wherein the ZIM CPU plays a critical role in achieving higher-order EPs, despite its small value. Direct evidence of this is provided by the simulated distributions of normalized magnetic fields $H_z/H_0$ (color map) and time-averaged Poynting vectors (arrows), computed using the finite-element software COMSOL Multiphysics, as presented in Fig. 2(d). The left, middle, and right panels correspond to the third-order lasing EP ($\alpha_1 \approx \alpha_2 \approx \alpha_3 \approx -i$), reflecting EP ($\alpha_1 = \left(0.5 + 0.5i\sqrt{3}\right) \times 10^{-3}$, $\alpha_2 = \left(-0.5 + 0.5i\sqrt{3}\right) \times 10^{-3}$, $\alpha_3 = -\sqrt{3}i \times 10^{-3}$) and absorbing EP ($\alpha_1 \approx \alpha_2 \approx \alpha_3 \approx i$), respectively, as marked by stars in Fig. 2(c). The incidence for all the three cases is identical, with complex magnetic-field amplitudes given by $H_0 e^{i2\pi/3}$, $H_0 e^{i\pi/3}$, and $H_0$ for the three channels, according to their identical eigenvectors $\psi = \left(e^{i2\pi/3}, e^{i\pi/3}, 1\right)^T$ (see Supplemental Material [64]). At the lasing EP, strong outgoing waves are observed; at the reflecting EP, standing waves arise due to total reflection; at the absorbing EP, all incident waves are completely absorbed (see Supplemental Material [64]).

We note that the proposed model could be realized in practice. The zero or purely imaginary values of $\alpha_{1,2,3}$ correspond to a zero or purely imaginary permeability of non-Hermitian ZIMs. Such permeability values can be achieved through the photonic doping approach [53-61]. Specifically, when an epsilon-near-zero (ENZ) material, characterized by a near-zero permittivity, is doped with nonmagnetic dopants exhibiting loss/gain in permittivity, the resulting composite can be homogenized as an effective ZIM. This effective ZIM retains a near-zero permittivity while acquiring a dopant-dependent effective permeability [53]. By adjusting the electromagnetic and geometric materials of the dopants, the effective permeability can be engineered across a wide range of values, including the purely imaginary ones (see Supplemental Material [64]). Notably, a hybrid doping strategy—combining a lossless dielectric dopant with a lossy/gain dopant—enables independent control over the real and imaginary parts of the effective permeability. The lossless dopant can be designed to nullify the real part, while the lossy/gain dopant tailors the imaginary component, yielding purely imaginary effective permeability [57]. In practice, the ENZ materials have been realized in various frequencies from microwaves to terahertz, infrared, and visible spectra [48, 49, 53]. Gain dopants can be realized by doping dielectrics with fluorescent dyes, quantum wells or dots in the optical and infrared regimes [65], and via negative-resistance-integrated metamaterials at microwave frequencies [66]. Additionally, ZIMs with zero permeability can also be implemented using dielectric photonic crystals exhibiting Dirac-like cone dispersions [50-52]. Here, we adopt the photonic doping approach to construct the proposed three-channel non-Hermitian ZIM models. Figure 2(e) presents the simulation results for the third-order lasing, reflecting, and absorbing EPs by doping ENZ materials with circular lossy/gain dopants, showing good agreement with those obtained using the ideal non-Hermitian ZIM models in Fig. 2(d). The detailed implementation method and parameters are provided in Supplemental Material [64].

The second solution of third-order EPs [Eq. (4)] exhibits a mathematical structure analogous to that of the first solution [Eq. (3)]. Consequently, the conditions for the occurrence of the third-order lasing, reflecting, and absorbing EPs are similar, that is,

$$\alpha_3 = -\left(1 - \sqrt{3}\,L/w\right)i, \ \ \alpha_3 = \sqrt{3}i\,L/w, \text{ and } \ \alpha_3 = \left(1 + \sqrt{3}\,L/w\right)i. \tag{7}$$

Their physical mechanisms and optical phenomena are consistent with those described for the first solution in Fig. 2.

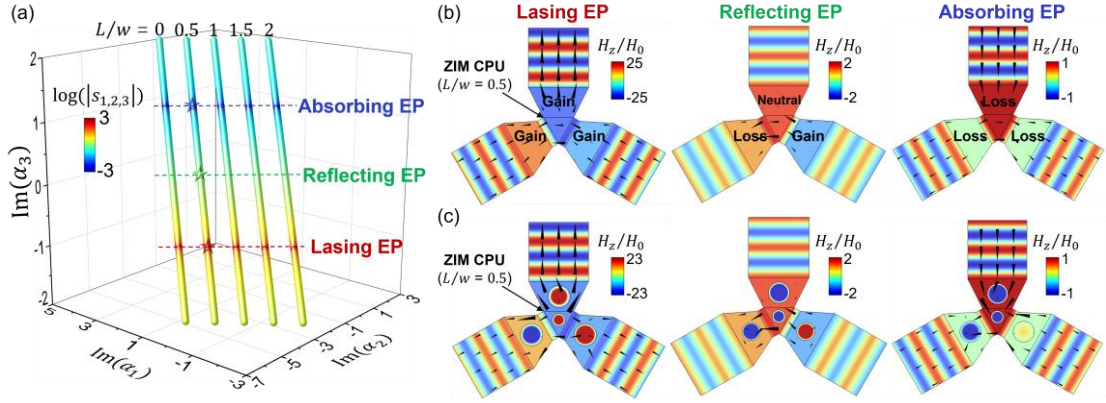

**FIG. 3.** (a) Evolution of the third-order EP of the third solution in the parameter space $\{\mathrm{Im}(\alpha_1), \mathrm{Im}(\alpha_2), \mathrm{Im}(\alpha_3)\}$ as $L/w$ increases, where $\alpha_1$, $\alpha_2$, and $\alpha_3$ are all considered purely imaginary. The color scale represents $\log|s_{1,2,3}|$. [(b) and (c)] Simulated distributions of normalized magnetic fields $H_z/H_0$ (color map) and time-averaged Poynting vectors (arrows) for the third-order lasing EP (left), reflecting EP (middle), and absorbing EP (right) when $L/w = 0.5$, as marked by stars in (a). Ideal non-Hermitian ZIM models and photonic doping models are considered in (b) and (c), respectively.

Next, we consider the third solution of the third-order EPs [Eq. (5)], which exhibits a distinctive property: it is independent of $L/w$. Notably, $L/w$-independent third-order lasing, reflecting, and absorbing EPs occur under the following conditions,

$$\alpha_3 = -i, \ \ \alpha_3 = 0, \text{ and } \ \alpha_3 = i. \tag{8}$$

Figure 3(a) shows the evolution of the third-order EPs of this solution in the parameter space $\{\mathrm{Im}(\alpha_1), \mathrm{Im}(\alpha_2), \mathrm{Im}(\alpha_3)\}$ as $L/w$ increases when $L = \lambda_0$, $A_C = \sqrt{3}L^2/4$, $A = 3\lambda_0^2$. The parameters $\alpha_1$, $\alpha_2$, and $\alpha_3$ are considered purely imaginary. The color scale represents $\log|s_{1,2,3}|$, showing the occurrence of $L/w$-independent third-order lasing, reflecting, and absorbing EPs. To further explore this case, we set $L/w = 0.5$. The normalized magnetic fields $H_z/H_0$ (color map) and time-averaged Poynting vectors (arrows) are shown in Figs. 3(b) and 3(c). The left, middle, and right panels correspond to the third-order lasing EP ($\alpha_1 = (-1 + 1/\sqrt{2})i$, $\alpha_2 = -(1 + 1/\sqrt{2})i$, $\alpha_3 = -i$), reflecting EP ($\alpha_1 = i/\sqrt{2}$, $\alpha_2 = -i/\sqrt{2}$, $\alpha_3 = 0$), and absorbing EP ($\alpha_1 = (1 + 1/\sqrt{2})i$, $\alpha_2 = (1 - 1/\sqrt{2})i$, $\alpha_3 = i$), respectively, as marked by stars in Fig. 3(a). For these cases, the incidence is identical, with complex magnetic-field amplitudes given by $i/\sqrt{2} H_0$, $-i/\sqrt{2} H_0$, and $H_0$ for the three channels, according to their identical eigenvectors $\psi = (i/\sqrt{2}, -i/\sqrt{2}, 1)^{\mathrm{T}}$ (see Supplemental Material [64]). The simulation results from the ideal non-Hermitian ZIM models [Fig. 3(b)] and the photonic doping models [Fig. 3(c)] (see Supplemental Material [64]) are in excellent agreement. Both reveal strong outgoing waves at the lasing EP, complete reflection at the reflecting EP, and perfect absorption at the absorbing EP. These results highlight the interesting wave behaviors associated with third-order EPs in the simplified three-channel non-Hermitian ZIM model. The emergence of EPs of higher order, exhibiting richer physical phenomena, can be anticipated with an increased number of input channels.

A prominent property of higher-order EPs is their ability to significantly enhance sensitivity to external perturbations, making them highly promising for sensing and detection applications [7, 8]. Here, we show a gas sensor using a three-channel non-Hermitian ZIM model operating near its third-order absorbing EP, where the output power undergoes a drastic change in response to perturbations in refractive index at the working frequency. Figure 4(a) illustrates the proposed gas sensor. The ZIM CPU is implemented using an ENZ material doped with a core-shell dopant, where a dielectric shell surrounds a vacuum core (see Supplemental Material [64]). In practical implementations, deep-subwavelength apertures can be created in the upper and lower plates within the vacuum core region. These apertures enable gases to pass through the vacuum core while preventing electromagnetic wave leakage (see Supplemental Material [64]). The gas flow induces a refractive index perturbation $\delta n$, leading to a deviation of the eigenvalues from their zero values at the absorbing EP.

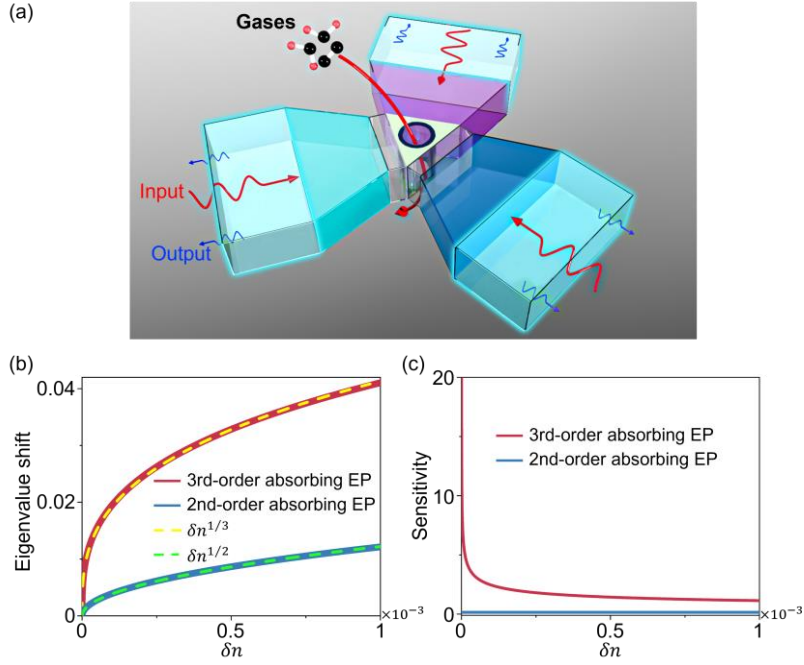

**FIG. 4.** (a) Schematic of a gas sensor employing a three-channel non-Hermitian ZIM model operating near its third-order absorbing EP. The ZIM CPU is implemented using an ENZ material doped with a core-shell dopant, where the vacuum core allows gases to pass through, inducing a refractive index perturbation $\delta n$. (b) Eigenvalue shift and (c) sensitivity versus the refractive index perturbation $\delta n$ in the dopant's core. The red (or blue) solid lines correspond to the three-channel (or two-channel) model operating near its third-order (or second-order) absorbing EP. The dashed lines in (b) represent the fitting curves.

It is interesting to point out that the gas sensor can be made of purely lossy ZIMs, when adopting the third solution for the third-order absorbing EP [Eq. (8)], where $\alpha_1 = i(1 + \sqrt{2}\,L/w)$, $\alpha_2 = i(1 - \sqrt{2}\,L/w)$ and $\alpha_3 = i$. When $L/w < \sqrt{2}/2$, the imaginary parts of $\alpha_1$, $\alpha_2$, and $\alpha_3$ are all positive, indicating purely lossy ZIMs. As an example, we set $L/w = 0.5$ to the gas sensor. The eigenvalue shift induced by the perturbation $\delta n$ is calculated and plotted as red solid lines in Fig. 4(b). As anticipated, the results align well with the theoretical $1/3$th power fitting curve (yellow dashed line). For comparison, the results for a two-channel model operating near its second-order absorbing EP are also shown, with the corresponding eigenvalue shift and $1/2$th power fitting curve represented by the blue solid and green dashed lines, respectively. We find that the $N$th-order absorbing EP exhibits an $N$th root dependence of eigenvalue shift on perturbations as that in coupled systems, and that higher-order absorbing EP provides a greater eigenvalue shift.

In practice, the eigenvalue shift is not the directly measurable quantity. Instead, the change in output power, induced by refractive index perturbation $\delta n$, can be experimentally observed. For a quantitative description, we define the total reflectance $R$ as ratio of the total output power to the total input power, which is calculated as $R = |s_n|^2$ when the input is carefully adjusted to fulfill the eigenvector associated with the shifted eigenvalue $s_n$. The sensitivity of the gas sensor is then defined as the change in total reflectance versus the refractive index perturbation, i.e., $\Delta R / \delta n$. Figure 4(c) shows the sensitivity as a function of the refractive index perturbation $\delta n$ in the dopant's core, in the vicinity of the absorbing EP, for both the three-channel and two-channel models, clearly demonstrating a significant enhancement in sensitivity near the third-order absorbing EP. This sensor enables the detection of low-concentration gases, as the variation in refractive index is directly proportional to gas concentration [67]. These findings validate the feasibility of ultrasensitive gas sensors operating near higher-order absorbing EPs.

Finally, it is noteworthy that unlike traditional coupled systems which exhibit higher-order EPs and are typically described by Hamiltonians [29-42], the proposed open scattering systems are analyzed using scattering matrices. It has been shown that well-defined EPs can also emerge in open scattering systems when two eigenvalues and their associated eigenvectors of the scattering matrices coalesce [4, 46]. Our findings reveal, for the first time, the presence of higher-order EPs and their associated intriguing wave behaviors in open scattering systems. Specifically, higher-order lasing, reflecting, and absorbing EPs are revealed. Superior to the conventional second-order absorbing EPs [14, 15], the higher-order absorbing EPs can give rise to enhanced sensing capabilities in a purely lossy ZIM system. In contrast to traditional coupled systems, which focus on perturbation-induced frequency splitting, our work emphasizes the drastic changes in output power induced by perturbations at a certain frequency, highlighting rich physical insights into higher-order EPs in open scattering systems.

The non-Hermitian ZIMs provide a viable platform for facilitating wave interference among multiple beams, supporting an almost arbitrary number of input channels. This flexibility enables, in principle, the realization of EPs of arbitrary order in open scattering systems, depending on the number of input channels. Although only third-order EPs are shown here, the emergence of EPs of higher order, which are expected to exhibit even richer physical phenomena, can be anticipated with an increased number of input channels. A four-channel model exhibiting fourth-order EPs is presented in Supplemental Material [64].

While multi-channel ZIM models have been studied previously, the EPs have never been discovered before [53-58, 62]. This is due to uniform fields across all ZIM regions, which inherently suppresses the emergence of EPs. Interestingly, our introduced air gaps, positioned between the ZIM CPU and input channels, disrupt the uniform field distribution and offer us additional degrees of freedom to tune the S-parameters, thereby enabling the EP realization. In practical implementations, these air gaps can be filled with dielectrics. The occurrence of EPs can be effectively controlled through adjusting the size of air gaps or the properties of the filling dielectrics, offering an effective approach for higher-order EP control.

In summary, we have revealed higher-order EPs through the interplay of loss and/or gain in ZIMs − factors traditionally considered detrimental in ZIM systems, revealing higher-order lasing, reflecting, and absorbing EPs. Notably, the higher-order absorbing EPs further give rise to an ultrasensitive gas sensor implemented using purely lossy ZIMs, highlighting its potential as an advanced platform for high-performance sensing applications.

**Acknowledgments.** National Natural Science Foundation of China (Grant Nos. 12374293, 11974010, 12274313, 12274314, 12274315, 12104191, 12474313, 12174188, 12474293); National Key Research and Development Program of China (2022YFA1404303); Natural Science Foundation of Jiangsu Province (Grant Nos. BK20221354, BK20221240, BK20233001); Suzhou Basic Research Project (Grant No. SJC2023003); Undergraduate Training Program for Innovation and Entrepreneurship, Soochow University (Grant No. 202310285027Z); The calculations of the behavior of exceptional points are partially supported by the Russian Science Foundation (Grant No. 23-72-00037).

**Data availability.** Data underlying the results presented in this paper are not publicly available at this time but may be obtained from the authors upon reasonable request.

**Disclosures.** The authors declare no conflicts of interest.

## References

[1] L. Feng, R. ElGanainy, and L. Ge, Non-Hermitian photonics based on parity-time symmetry, Nat. Photonics **11**, 752-762 (2017).

[2] R. El-Ganainy, K. G. Makris, M. Khajavikhan, Z. H. Musslimani, S. Rotter, and D. N. Christodoulides, Non-Hermitian physics and PT symmetry, Nat. Phys. **14**, 11 (2017).

[3] Y. Huang, Y. Shen, C. Min, S. Fan, and G. Veronis, Unidirectional reflectionless light propagation at exceptional points, Nanophotonics **6**, 977-996 (2017).

[4] M. Miri, and A. Alù, Exceptional points in optics and photonics, Science **363**, eaar7709 (2019).

[5] S. K. Özdemir, S. Rotter, F. Nori, and L. Yang, Parity-time symmetry and exceptional points in photonics, Nat. Mater. **18**, 783-798 (2019).

[6] B. Qi, H. Z. Chen, L. Ge, P. Berini, and R. M. Ma, Parity-time symmetry synthetic lasers: Physics and devices, Adv. Opt. Mater. **7**, 1900694 (2019).

[7] J. Wiersig, Review of exceptional point-based sensors, Photonics Res. **8**, 1457 (2020).

[8] M. De Carlo, F. De Leonardis, R. A. Soref, L. Colatorti, and V. M. N. Passaro, Non-Hermitian sensing in photonics and electronics: A review, Sensors **22**, 3977 (2022).

[9] K. Ding, C. Fang, and G. Ma, Non-Hermitian topology and exceptional-point geometries, Nat. Rev. Phys. **4**, 745-760 (2022).

[10] A. Li, H. Wei, M. Cotrufo, W. Chen, S. Mann, X. Ni, B. Xu, J. Chen, J. Wang, S. Fan, C. Qiu, A. Alù, and L. Chen, Exceptional points and non-Hermitian photonics at the nanoscale, Nat. Nanotechnol. **18**, 706-720 (2023).

[11] P. Miao, Z. Zhang, J. Sun, W. Walasik, S. Longhi, N. M. Litchinitser, and L. Feng, Orbital angular momentum microlaser, Science **353**, 464-467 (2016).

[12] J. Zhang, B. Peng, K. Özdemir, K. Pichler, D. O. Krimer, G. Zhao, F. Nori, Y. Liu, S. Rotter, and L. Yang, A phonon laser operating at an exceptional point, Nat. Photonics **12**, 479-484 (2018).

[13] B. Peng, S. K. Özdemir, M. Liertzer, W. Chen, J. Kramer, H. Yılmaz, J. Wiersig, S. Rotter, and L. Yang, Chiral modes and directional lasing at exceptional points, Proc. Natl. Acad. Sci. U. S. A. **113**, 6845-6850 (2016).

[14] W. R. Sweeney, C. W. Hsu, S. Rotter, and A. D. Stone, Perfectly absorbing exceptional points and chiral absorbers, Phys. Rev. Lett. **122**, 093901 (2019).

[15] C. Wang, W. R. Sweeney, A. D. Stone, and L. Yang, Coherent perfect absorption at an exceptional point, Science **373**, 1261-1265 (2021).

[16] D. L. Sounas, R. Fleury, and A. Alù, Unidirectional cloaking based on metasurfaces with balanced loss and gain, Phys. Rev. Appl. **4**, 014005 (2015).

[17] J. Luo, J. Li, and Y. Lai, Electromagnetic impurity-immunity induced by parity-time symmetry, Phys. Rev. X. **8**, 031035 (2018).

[18] C. Liu, D. Yan, B. Sun, Y. Xu, F. Cao, L. Gao, and J. Luo, Low-gain generalized PT symmetry for electromagnetic impurity-immunity via non-Hermitian doped zero-index materials, Photonics Res. **12**, 2424 (2024).

[19] J. Wiersig, Enhancing the sensitivity of frequency and energy splitting detection by using exceptional points: Application to microcavity sensors for single-particle detection, Phys. Rev. Lett. **112**, 203901 (2014).

[20] Z. Liu, J. Zhang, S. K. Ozdemir, B. Peng, H. Jing, X. Lü, C. Li, L. Yang, F. Nori, and Y. Liu,

Metrology with PT-symmetric cavities: Enhanced sensitivity near the PT-phase transition, Phys. Rev. Lett. **117**, 110802 (2016).

[21] J. Wiersig, Sensors operating at exceptional points: General theory, Phys. Rev. A **93**, 033809 (2016).

[22] W. Chen, S. K. Ozdemir, G. Zhao, J. Wiersig, and L. Yang, Exceptional points enhance sensing in an optical microcavity, Nature **548**, 192-196 (2017).

[23] H. Zhao, Z. Chen, R. Zhao, and L. Feng, Exceptional point engineered glass slide for microscopic thermal mapping, Nat. Commun. **9**, 1764 (2018).

[24] M. P. Hokmabadi, A. Schumer, D. N. Christodoulides, and M. Khajavikhan, Non-Hermitian ring laser gyroscopes with enhanced Sagnac sensitivity, Nature **576**, 70-74 (2019).

[25] Y. Lai, Y. Lu, M. Suh, Z. Yuan, and K. Vahala, Observation of the exceptional-point-enhanced Sagnac effect, Nature **576**, 65-69 (2019).

[26] Q. Zhong, J. Ren, M. Khajavikhan, D. N. Christodoulides, K. Özdemir, and R. El-Ganainy, Sensing with exceptional surfaces in order to combine sensitivity with robustness, Phys. Rev. Lett. **122**, 153902 (2019).

[27] Z. Dong, Z. Li, F. Yang, C. Qiu, and J. S. Ho, Sensitive readout of implantable microsensors using a wireless system locked to an exceptional point, Nat. Electron. **2**, 335-342 (2019).

[28] J. Park, A. Ndao, W. Cai, L. Hsu, A. Kodigala, T. Lepetit, Y. Lo, and B. Kanté, Symmetry-breaking-induced plasmonic exceptional points and nanoscale sensing, Nat. Phys. **16**, 462-468 (2020).

[29] K. Ding, G. Ma, M. Xiao, Z. Q. Zhang, and C. T. Chan, Emergence, coalescence, and topological properties of multiple exceptional points and their experimental realization, Phys. Rev. X. **6**, 021007 (2016).

[30] Z. Lin, A. Pick, M. Lončar, and A. W. Rodriguez, Enhanced spontaneous emission at third-order Dirac exceptional points in inverse-designed photonic crystals, Phys. Rev. Lett. **117**, 107402 (2016).

[31] H. Hodaei, A. U. Hassan, S. Wittek, H. Garcia-Gracia, R. El-Ganainy, D. N. Christodoulides, and M. Khajavikhan, Enhanced sensitivity at higher-order exceptional points, Nature **548**, 187-191 (2017).

[32] J. Kullig, D. Grom, S. Klembt, and J. Wiersig, Higher-order exceptional points in waveguide-coupled microcavities: perturbation induced frequency splitting and mode patterns, Photonics Res. **11**, A54 (2023).

[33] Y. Feng, Y. Wang, Z. Li, and T. Li, Enhanced sensing and broadened absorption with higher-order scattering zeros, Opt. Express **32**, 32283 (2024).

[34] J. Kullig, and J. Wiersig, High-order exceptional points of counterpropagating waves in weakly deformed microdisk cavities, Phys. Rev. A **100**, 043837 (2019).

[35] Z. Xiao, H. Li, T. Kottos, and A. Alù, Enhanced sensing and nondegraded thermal noise performance based on PT-symmetric electronic circuits with a sixth-order exceptional point, Phys. Rev. Lett. **123**, 213901 (2019).

[36] C. Zeng, Y. Sun, G. Li, Y. Li, H. Jiang, Y. Yang, and H. Chen, Enhanced sensitivity at high-order exceptional points in a passive wireless sensing system, Opt. Express **27**, 27562 (2019).

[37] M. Sakhdari, M. Hajizadegan, and P. Chen, Robust extended-range wireless power transfer using a higher-order PT-symmetric platform, Phys. Rev. Res. **2**, 013152 (2020).

[38] T. Chen, D. Zou, Z. Zhou, R. Wang, Y. Feng, H. Sun, and X. Zhang, Ultra-sensitivity in reconstructed exceptional systems, Natl. Sci. Rev. **11**, nwae278 (2024).

[39] Q. Zhong, J. Kou, K. Özdemir, and R. El-Ganainy, Hierarchical construction of higher-order exceptional points, Phys. Rev. Lett. **125**, 203602 (2020).

[40] S. Wang, B. Hou, W. Lu, Y. Chen, Z. Q. Zhang, and C. T. Chan, Arbitrary order exceptional point induced by photonic spin-orbit interaction in coupled resonators, Nat. Commun. **10**, 832 (2019).

[41] Y. Zhang, S. Xia, X. Zhao, L. Qin, X. Feng, W. Qi, Y. Jiang, H. Lu, D. Song, L. Tang, Z. Zhu, W. Liu, and Y. Liu, Symmetry-protected third-order exceptional points in staggered flatband rhombic lattices, Photonics Res. **11**, 225 (2023).

[42] N. Wang, B. Hong, and G. P. Wang, Higher-order exceptional points and enhanced polarization sensitivity in anisotropic photonic time-Floquet crystals, Opt. Express **32**, 40092 (2024).

[43] A. Figotin, and I. Vitebskiy, Oblique frozen modes in periodic layered media, Phys. Rev. E **68**, 036609 (2003).

[44] A. Figotin, and I. Vitebskiy, Frozen light in photonic crystals with degenerate band edge, Phys. Rev. E **74**, 066613 (2006).

[45] M. Y. Nada, M. A. K. Othman, and F. Capolino, Theory of coupled resonator optical waveguides exhibiting high-order exceptional points of degeneracy, Phys. Rev. B **96**, 184304 (2017).

[46] Y. D. Chong, L. Ge, and A. D. Stone, PT-symmetry breaking and laser-absorber modes in optical scattering systems, Phys. Rev. Lett. **106**, 093902 (2011).

[47] J. Erb, N. Shaibe, R. Calvo, D. P. Lathrop, Jr. T. M. Antonsen, T. Kottos, and S. M. Anlage, Novel topology and manipulation of scattering singularities in complex non-Hermitian systems, arXiv: 2411.01069v2.

[48] I. Liberal, and N. Engheta, Near-zero refractive index photonics, Nat. Photonics **11**, 149-158 (2017).

[49] X. Niu, X. Hu, S. Chu, and Q. Gong, Epsilon-near-zero photonics: A new platform for integrated devices, Adv. Opt. Mater. **6**, 1701292 (2018).

[50] Y. Li, C. T. Chan, and E. Mazur, Dirac-like cone-based electromagnetic zero-index metamaterials, Light-Sci. Appl. **10**, 203 (2021).

[51] X. Huang, Y. Lai, Z. H. Hang, H. Zheng, and C. T. Chan, Dirac cones induced by accidental degeneracy in photonic crystals and zero-refractive-index materials, Nat. Mater. **10**, 582-586 (2011).

[52] J. Luo, and Y. Lai, Hermitian and non-Hermitian Dirac-like cones in photonic and phononic structures, Front. Phys.-Lausanne **10**, 845624 (2022).

[53] I. Liberal, A. M. Mahmoud, Y. Li, B. Edwards, and N. Engheta, Photonic doping of epsilon-near-zero media, Science **355**, 1058-1062 (2017).

[54] J. Luo, B. Liu, Z. H. Hang, and Y. Lai, Coherent perfect absorption via photonic doping of zero-index media, Laser Photon. Rev. **12**, 1800001 (2018).

[55] D. Yan, R. Mei, M. Li, Z. Ma, Z. H. Hang, and J. Luo, Controlling coherent perfect absorption via long-range connectivity of defects in three-dimensional zero-index media, Nanophotonics **12**, 4205-4214 (2023).

[56] Z. Zhou, and Y. Li, N-port equal/unequal-split power dividers using epsilon-near-zero metamaterials, IEEE Trans. Microw. Theory Tech. **69**, 1529-1537 (2021).

[57] W. Ji, D. Wang, S. Li, Y. Shang, W. Xiong, L. Zhang, and J. Luo, Photonic-doped epsilon-near-zero media for coherent perfect absorption, Appl. Phys. A **125**, 129 (2019).

[58] Y. Wang, J. Lin, and P. Xu, Transmission-reflection decoupling of non-Hermitian photonic doping epsilon-near-zero media, Front. Phys. **19**, 33206 (2024).

[59] W. Yan, H. Li, X. Qin, P. Li, P. Fu, K. Li, and Y. Li, Fano resonance in epsilon-near-zero media, Phys. Rev. Lett. **133**, 256402 (2024).

[60] W. Yan, Z. Zhou, H. Li, and Y. Li, Transmission-type photonic doping for high-efficiency epsilon-near-zero supercoupling, Nat. Commun. **14**, 6154 (2023).

[61] M. Coppolaro, M. Moccia, G. Castaldi, N. Engheta, and V. Galdi, Non-Hermitian doping of epsilon-near-zero media, Proc. Natl. Acad. Sci. U. S. A. **117**, 13921-13928 (2020).

[62] I. Liberal, and N. Engheta, Multiqubit subradiant states in N-port waveguide devices: $\varepsilon$-and-$\mu$-near-zero hubs and nonreciprocal circulators, Phys. Rev. A **97**, 022309 (2018).

[63] D. M. Pozar, *Microwave Engineering* (John Wiley & Sons, Inc., 2012), 4 ed.

[64] See Supplemental Material.

[65] O. Hess, J. B. Pendry, S. A. Maier, R. F. Oulton, J. M. Hamm, and K. L. Tsakmakidis, Active nanoplasmonic metamaterials, Nat. Mater. **11**, 573-584 (2012).

[66] C. Qian, Y. Yang, Y. Hua, C. Wang, X. Lin, T. Cai, D. Ye, E. Li, I. Kaminer, and H. Chen, Breaking the fundamental scattering limit with gain metasurfaces, Nat. Commun. **13**, 4383 (2022).

[67] M. Njegovec, and D. Donlagic, A fiber-optic gas sensor and method for the measurement of refractive index dispersion in NIR, Sensors **20**, 3717 (2020).

# Supplemental Material

## 1. Scattering matrix for *N*-port model

Figure S1 shows the top-down view of the studied non-Hermitian zero-index material (ZIM) model. It consists of a regular polygon-shaped ZIM "central processing unit (CPU)" with side length $L$, connected to $N$ channels of identical width $w$, each separated by an air gap $g_n$ in the $n$th channel. Each channel comprises a hollow straight region, enabling wave input and output, and a trapezoid region filled with a unique ZIM for seamless coupling with the ZIM CPU, as illustrated in the right panel.

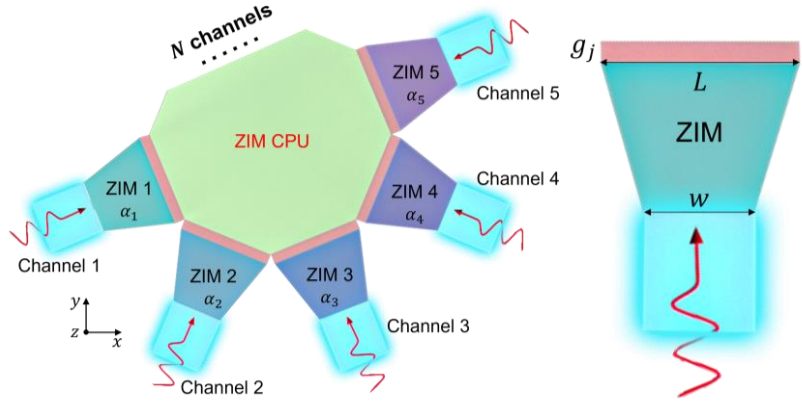

**Figure S1.** Left: Top-down view of the non-Hermitian ZIM model for higher-order EPs. Right: The enlarged view of one channel.

We consider plane waves of transverse magnetic (TM) polarization with out-of-plane magnetic fields (along the $z$ direction). To derive the explicit expression of the scattering matrix describing this $N$-channel model, we first assume that only the $n$th channel is excited by a wave characterized by a complex magnetic-field amplitude $H_n$. Then, the magnetic field within $m$th channel can be expressed as,

$$\mathbf{H}_m = H_n\big(\delta_{mn}e^{i\mathbf{k}_0\cdot\mathbf{r}} + S_{mn}e^{-i\mathbf{k}_0\cdot\mathbf{r}}\big)\mathbf{e}_z, \tag{S1}$$

where $S_{mn}$ is the S-parameter, defined as the ratio of the output magnetic field within the $m$th channel to the input magnetic field within the $n$th channel. $\mathbf{k}_0$ is the wave vector in the hollow region. Here, the time harmonic term $e^{-i\omega t}$ is omitted. According to the Ampère-Maxwell equation,

$$\mathbf{E}_m = \frac{i}{\omega\varepsilon}\nabla\times\mathbf{H}_m, \tag{S2}$$

the electric field within $m$th channel is derived as:

$$\mathbf{E}_m = Z_0 H_n \big( \delta_{mn} e^{i\mathbf{k}_0 \cdot \mathbf{r}} - S_{mn} e^{-i\mathbf{k}_0 \cdot \mathbf{r}} \big) \mathbf{e}_\tau, \tag{S3}$$

where $Z_0$ is the impedance of vacuum; $\omega$ is the angular frequency; $\mathbf{e}_\tau$ is the unit vector along the tangential direction of the channel. Similarly, the magnetic and electric fields within $m$th air gap are expressed as,

$$\mathbf{H}_{g,m} = H_m \big( a_m e^{i\mathbf{k}_0 \cdot \mathbf{r}} + b_m e^{-i\mathbf{k}_0 \cdot \mathbf{r}} \big) \mathbf{e}_z, \tag{S4}$$

$$\text{and } \mathbf{E}_{g,m} = Z_0 H_m \big( a_m e^{i\mathbf{k}_0 \cdot \mathbf{r}} - b_m e^{-i\mathbf{k}_0 \cdot \mathbf{r}} \big) \mathbf{e}_\tau, \tag{S5}$$

where $H_m a_m$ and $H_m b_m$ represent the complex magnetic-field amplitudes of the forward and backward waves within the air gap, respectively. Considering the uniform magnetic field within the ZIM, and the boundary continuity condition at the ZIM-gap interface for the $m$th channel, we have,

$$\delta_{mn} + S_{mn} = a_m + b_m. \tag{S6}$$

Then, considering the uniform magnetic field within ZIM CPU and the boundary continuity condition at the gap-ZIM CPU interface, the magnetic fields within the $m$th and the $p$th channels can be connected via,

$$a_m e^{ik_0 g_m} + b_m e^{-ik_0 g_m} = a_p e^{ik_0 g_p} + b_p e^{-ik_0 g_p}. \tag{S7}$$

Applying the Faraday-Maxwell equation to the $m$th ZIM and the ZIM CPU, we obtain, respectively,

$$-w(\delta_{mn} - S_{mn}) + L(a_m - b_m) = ik_0 \mu_m A(a_m + b_m), \tag{S8}$$

$$\text{and } -\sum_{p=1}^{N} L\big( a_p e^{ik_0 g_p} - b_p e^{-ik_0 g_p} \big) = ik_0 \mu_C A_C \big( a_m e^{ik_0 g_m} + b_m e^{-ik_0 g_m} \big), \tag{S9}$$

where $A$ and $A_C$ are the area of the ZIM CPU and the $m$th ZIM within the channel, respectively. Based on Eqs. (S6)-(S9), the S-parameter can be obtained as,

$$S_{mn} = \frac{-2L/w}{[(1 - i\alpha_m)\sin(k_0 g_m) + i(L/w)\cos(k_0 g_m)][(1 - i\alpha_n)\sin(k_0 g_n) + i(L/w)\cos(k_0 g_n)]\gamma}$$
$$+ \delta_{mn} \frac{(1 + i\alpha_m)\sin(k_0 g_m) - i(L/w)\cos(k_0 g_m)}{(1 - i\alpha_m)\sin(k_0 g_m) + i(L/w)\cos(k_0 g_m)}, \tag{S10}$$

where $\alpha_{m(n)} = \frac{k_0 A \mu_{m(n)}}{w}$, $\gamma = \sum_{n=1}^{N} \frac{(1 - i\alpha_n)\cos(k_0 g_n) - (L/w)i\sin(k_0 g_n)}{(L/w)\cos(k_0 g_n) - i(1 - i\alpha_n)\sin(k_0 g_n)} - i\alpha_C w/L$ with $\alpha_C = \frac{k_0 A_C \mu_C}{w}$.

## 2. Solutions for higher-order EPs

In the $N$-channel model, the eigen polynomial of the scattering matrix $S$ can be expanded as,

$$|sI - S| = s^N - \text{Tr}(S)s^{N-1} + \cdots + (-1)^j \sum_l |S|_l^{(j)} s^{N-j} + \cdots + (-1)^N |S|, \quad (S11)$$

where $\sum_l |S|_l^{(j)} s^{N-j}$ represents the sum of all the $j$th order principal minors of the scattering matrix; $I$ is the unitary matrix; $s$ is the eigenvalue.

Assuming that the model exhibits an $N$th-order EP, then the eigen equation $|sI - S| = 0$ would possess $N$ identical complex roots, which can be expressed as $s = \frac{\text{Tr}(S)}{N}$ according to Vieta's theorem. In this case, the eigen polynomial in Eq. (S11) can be rewritten as $\left(s - \frac{\text{Tr}(S)}{N}\right)^N$, which can be expanded using the binomial theorem as

$$|sI - S| = s^N - \text{Tr}(S)s^{N-1} + \cdots + (-1)^j \frac{N!}{j!(N-j)!} \left[\frac{\text{Tr}(S)}{N}\right]^j s^{N-j} + \cdots + (-1)^N \left[\frac{\text{Tr}(S)}{N}\right]^N. \quad (S12)$$

By subtracting Eq. (S12) from Eq. (S11), we obtain

$$\sum_{j=2}^{N} (-1)^j \left\{ \sum_l |S|_l^{(j)} - \frac{N!}{j!(N-j)!} \left[\frac{\text{Tr}(S)}{N}\right]^j \right\} s^{N-j} \equiv 0. \quad (S13)$$

Equation (S13) indicates that the presence of the $N$th-order EP requires the satisfaction of the following condition:

$$\sum_l |S|_l^{(j)} - \frac{N!}{j!(N-j)!} \left[\frac{\text{Tr}(S)}{N}\right]^j = 0 \quad (j = 2,3,\ldots,N). \quad (S14)$$

Equation (S14) comprises $N - 1$ equations based on which the $N$th-order degenerate eigenvalue $\frac{\text{Tr}(S)}{N}$ can be identified. However, the $N$th-order degenerate eigenvalue is not necessarily the $N$th-order EP unless the following condition is further satisfied:

$$\text{rank}(|sI - S|) = N - 1. \quad (S15)$$

Equation (S15) stems from the requirement of coalescence of eigenvectors at EP. Based on Eqs. (S10), (S14), and (S15), EP of arbitrary order, in principle, can be identified for an $N$-channel non-Hermitian ZIM model.

For a simplified three-channel model, Eq. (S14) gives rise to

$$\left(\frac{S_{11}+S_{22}+S_{33}}{3}\right)^3 - \begin{vmatrix} S_{11} & S_{12} & S_{13} \\ S_{21} & S_{22} & S_{23} \\ S_{31} & S_{32} & S_{33} \end{vmatrix} = 0, \quad (S16)$$

$$\text{and } 3\left(\frac{S_{11}+S_{22}+S_{33}}{3}\right)^2 - \begin{vmatrix} S_{11} & S_{12} \\ S_{21} & S_{22} \end{vmatrix} - \begin{vmatrix} S_{11} & S_{13} \\ S_{31} & S_{33} \end{vmatrix} - \begin{vmatrix} S_{22} & S_{23} \\ S_{32} & S_{33} \end{vmatrix} = 0, \quad (S17)$$

By substituting the scattering matrix in Eq. (S10) into Eqs. (S16) and (S17), we can determine the three-channel non-Hermitian ZIM model that exhibits $N$th-order EP provided that the condition $\text{rank}(|sI - S|) = 2$ is satisfied, as required by Eq. (S15).

In the main text, we set $\mu_C \approx 0$, $g_1 = g_2 = \lambda_0/4$ and $g_3 = 0$ for the three-channel model. Under this circumstance, we can obtain three solutions for the third-order EPs based on Eqs. (S16) and (S17), that is, Eqs. (3)-(5) in the main text. A notable coincidence is that the eigenvectors associated with all eigenvalues for each solution are identical. For the first solution of the third-order EPs [Eq. (3)], the eigenvector is $\psi = \left(e^{i2\pi/3}, e^{i\pi/3}, 1\right)^{\mathrm{T}}$; for the second solution, the eigenvector is $\psi = \left(e^{i4\pi/3}, e^{i5\pi/3}, 1\right)^{\mathrm{T}}$; for the third solution, the eigenvector is $\left(i/\sqrt{2}, -i/\sqrt{2}, 1\right)^{\mathrm{T}}$.

### 3. Third-order reflecting EP

In this section, we investigate the newly discovered reflecting EP in detail. Let us consider the three-channel non-Hermitian ZIM model, which is characterized by a $3 \times 3$ scattering matrix $S$. The emergence of reflecting EPs demands that the outgoing power within each channel equals the input power. This leads to the condition for S-parameter $S_{m1}\psi_1 + S_{m2}\psi_2 + S_{m3}\psi_3 = \psi_m$ with $m = 1,2,3$, or the condition for eigenvalues $s_{1,2,3} = 1$. Here, $S_{m1}$ is $m$-th row and 1-st column element (S-parameter) of the scattering matrix $S$. $\psi_m$ is the $m$-th element of the eigenvector, corresponding to the incident field within the $m$-th channel.

Specifically, in the main text, we have shown that the third-order reflecting EP associated with the first solution of third-order EP [Eqs. (3) and (6)] appears when $\alpha_1 = \alpha_3 + \left(1 + 3i\sqrt{3}\right) L/(2w)$, $\alpha_2 = \alpha_3 - \left(1 - 3i\sqrt{3}\right) L/(2w)$, and $\alpha_3 = -\sqrt{3}iL/w$. Under this circumstance, the scattering matrix is,

$$S_{\mathrm{REP1}} = \begin{pmatrix} 1 - (1+\sqrt{3}i)(L/w - i)L/w & -2L^2/w^2 & (2 + (\sqrt{3} + i)L/w)iL/w \\ -2L^2/w^2 & 1 + (i + \sqrt{3})(L/w + i)iL/w & (2i + (\sqrt{3}i + 1)L/w)L/w \\ (2 + (\sqrt{3} + i)L/w)iL/w & (2i + (\sqrt{3}i + 1)L/w)L/w & 1 + 2L/w(L/w + \sqrt{3}) \end{pmatrix}. \text{(S18)}$$

It can be shown that the matrix $S_{\mathrm{REP1}}$ has three identical eigenvalues $s_{1,2,3} = 1$ and three corresponding coalesced eigenvectors $\psi = \left(e^{i2\pi/3}, e^{i\pi/3}, 1\right)^{\mathrm{T}}$, indicating the emergence of third-order reflecting EP. A similar behavior can be observed from the second solution [Eq. (4)].

For the third solution of third-order EP [Eqs. (5) and (8)], the reflecting EP occurs when $\alpha_1 = \alpha_3 + i\sqrt{2}L/w$, $\alpha_2 = \alpha_3 - i\sqrt{2}L/w$, and $\alpha_3 = 0$, corresponding to a scattering matrix:

$$S_{\text{REP3}} = \begin{pmatrix} \left(-1 + \sqrt{2}L/w\right)^2 & -2L^2/w^2 & 2iL/w\left(1 - \sqrt{2}L/w\right) \\ -2L^2/w^2 & \left(1 + \sqrt{2}L/w\right)^2 & 2iL/w\left(1 + \sqrt{2}L/w\right) \\ 2iL/w\left(1 - \sqrt{2}L/w\right) & 2iL/w\left(1 + \sqrt{2}L/w\right) & 1 - 4L^2/w^2 \end{pmatrix}. \quad (S19)$$

It can be proved that the matrix $S_{\text{REP3}}$ has three identical eigenvalues $s_{1,2,3} = 1$ and three corresponding coalesced eigenvectors $\psi = \left(i/\sqrt{2}, -i/\sqrt{2}, 1\right)^{\text{T}}$, indicating the emergence of third-order reflecting EP.

The above results show that the reflecting EP does not suggest that all reflection coefficients are unity (i.e., $S_{mm} = 1$). Rather, it requires the relation $S_{m1}\psi_1 + S_{m2}\psi_2 + S_{m3}\psi_3 = \psi_m$ for $m = 1,2,3$. When we carefully examine the energy flow within the non-Hermitian ZIM model operating at the reflecting EP, we can see energy exchange among different non-Hermitian ZIM regions. From the third-order reflecting EP model studied in Fig. 3, a clear energy flow from the gain non-Hermitian ZIM to the lossy one can be seen.

## 4. Fourth-order EPs

Increasing the number of channels enables, in principle, the realization of EPs of arbitrary order. Utilizing Eqs. (S10), (S14), and (S15), we identify a four-channel model that exhibits fourth-order EPs. The relevant parameters for this model, facilitating the emergence of fourth-order lasing, reflecting, and absorbing EPs when $L/w = 0.5$, are presented in Table 1.

**Table 1.** The relevant parameters for the four-channel model exhibiting fourth-order EPs.

| Parameters | Lasing EP | Reflecting EP | Absorbing EP |
|---|---|---|---|
| $\alpha_1$ | $-0.4019 - 0.6240i$ | $-0.4019 - 0.3562i$ | $-0.4019 + 0.6438i$ |
| $\alpha_2$ | $-0.5253i$ | $-0.4549i$ | $0.5451i$ |
| $\alpha_3$ | $0.4019 + 0.6240i$ | $0.4019 - 0.3562i$ | $0.4019 + 0.6438i$ |
| $\alpha_4$ | $-2.1474i$ | $1.1672i$ | $2.16721i$ |

For the lasing EP, the eigenvector is $\psi = (0.31 - 0.35i, 0.55, 0.31 + 0.35i, -0.5i)^{\text{T}}$; for the reflecting EP, the eigenvector is $\psi = (0.31 + 0.35i, 0.55, 0.31 - 0.5i, 0.5i)^{\text{T}}$; for the absorbing EP, the eigenvector is $\psi = (0.31 + 0.35i, 0.55, 0.31 - 0.35i, 0.5i)^{\mathrm{T}}$.

## 5. Role of air gaps in higher-order EP realization

In open scattering systems, higher-order EPs do not emerge in every scattering matrix formalism. To achieve higher-order EPs, it is essential to have sufficient degrees of freedom to freely manipulate the elements of the scattering matrix (i.e., S parameters) to conform to the desired formalism. The air gaps in our system are purposefully introduced to provide this degree of freedom, as elaborated as follows.

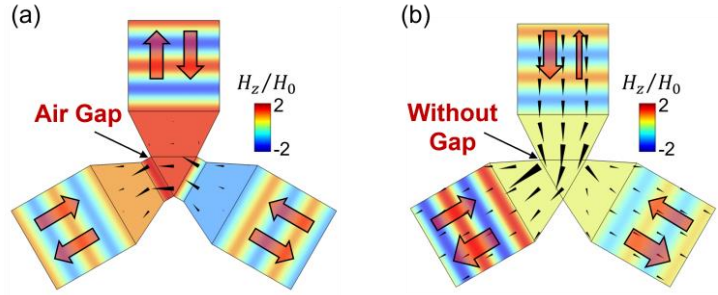

**Figure S2.** (a) Normalized magnetic fields $H_z/H_0$ (color map) and time-averaged Poynting vectors (arrows) for the reflecting EP model studied in Fig. 3(b) in the main text. Distinct values of magnetic fields are observed across different ZIM regions. (b) Air gaps have been removed. The magnetic fields across all ZIM regions become uniform (color map), and net time-averaged Poynting vectors occur within channels (arrows), indicating the disappearance of the reflecting EP.

Let us first consider a two-dimensional ZIM situated on the $xy$ plane illuminated by a TM polarized wave with magnetic field oriented along the $z$ direction. Given that this ZIM has a near-zero permittivity (i.e., $\varepsilon_{\mathrm{ZIM}} \to 0$), according to the Ampère-Maxwell equation $\nabla \times \mathrm{H} = -i\omega\varepsilon_{\mathrm{ZIM}}\mathrm{E}$, when $\varepsilon_{\mathrm{ZIM}} \to 0$, we have $\frac{\partial H_z}{\partial x} = \frac{\partial H_z}{\partial y} \to 0$. This implies that the magnetic field within the ZIM region is uniformly distributed. In the absence of air gaps, the magnetic field across all ZIM regions is uniform, which imposes constraints on the correlations among the S-parameters. Considering the boundary continuity condition of magnetic field within each channel, we obtain the relation $1 + S_{nn} = S_{mn}$ $(m \neq n)$. In this situation, the eigenvalues of the scattering matrix $S$ are $s_1 = s_2 = \cdots = s_{N-1} = -1$ and $s_N = N - 1 + \sum_{n=1}^{N} S_{nn}$.

Although $N-1$ eigenvalues are degenerate, their eigenvectors remain orthogonal, which prevents the emergence of EPs.

Interestingly, we find that the induction of air gaps can disrupt the uniformity of the magnetic field across all ZIM regions. While the magnetic field within each ZIM region remains constant, this disruption breaks the condition $1 + S_{nn} = S_{mn}$. Consequently, the air gaps offer us additional degrees of freedom to tune the S-parameters, thereby enabling the realization of higher-order EPs.

To verify this, we present a numerical example in Fig. S2. Here, we take the reflecting EP model studied in Fig. 3(b) in the main text as an example. Figure S2(a) shows the simulated normalized magnetic fields $H_z/H_0$ (color map) and time-averaged Poynting vectors (arrows) for the reflecting EP model. Distinct values of magnetic field are observed in different ZIM regions. However, when the air gaps are removed [Fig. S2(b)], the magnetic field across all ZIM regions becomes uniform, and net time-averaged Poynting vectors appear within the channels (arrows), indicating the disappearance of the reflecting EP. These results underscore the crucial role of air gaps in disrupting the uniformity of magnetic field across all ZIM regions and in facilitating the realization of higher-order EPs.

## 6. Photonic doping approach for non-Hermitian ZIMs

In Figs. 2 and 3 in the main text, we adopt the photonic doping approach [1] for realizing the desired non-Hermitian ZIMs. In this section, we provide a detailed elaboration of this approach. We show that through doping an epsilon-near-zero (ENZ) material with appropriate dopants, a wide range of effective permeability values could be obtained. To illustrate this, let us consider a nonmagnetic circular dopant (radius $r$, relative permittivity $\varepsilon$, relative permeability $\mu = 1$) embedded inside an ENZ material (relative permittivity $\varepsilon_{\text{ENZ}} \approx 0$, relative permeability $\mu_{\text{ENZ}} = 1$). According to the photonic doping approach, under the illumination of TM polarization, the effective relative permittivity $\varepsilon_{\text{eff}}$ of the doped ENZ material remains near-zero, i.e., $\varepsilon_{\text{eff}} \approx 0$. Meanwhile, the effective relative permeability $\mu_{\text{eff}}$ depends on the dopant and can be expressed as,

$$\mu_{\text{eff}} = \frac{1}{A_{\text{total}}} \left[ A_{\text{total}} - \pi r^2 + \frac{2\pi r J_1(\sqrt{\varepsilon}k_0 r)}{\sqrt{\varepsilon}k_0 J_0(\sqrt{\varepsilon}k_0 r)} \right], \tag{S20}$$

where $A_{\text{total}}$ is the total area of the ENZ material and the dopant. $J_0(\dots)$ and $J_1(\dots)$ are the 0-th and 1-st order Bessel functions, respectively.

Figure S3 presents a numerical example, showing the real part $\text{Re}(\mu_{\text{eff}})$ and imaginary part $\text{Im}(\mu_{\text{eff}})$ of $\mu_{\text{eff}}$ as functions of $\text{Re}(\varepsilon)$ and $\text{Im}(\varepsilon)$ of the dopant based on Eq. (S20). It is seen that a wide range of values of $\text{Im}(\mu_{\text{eff}})$ can be obtained. In practice, the dopant's radius $r$ can also be adjusted as an additional degree of freedom for engineering $\mu_{\text{eff}}$. In this manner, we can design non-Hermitian ZIMs with desired imaginary permeability simply using available dielectrics with loss or gain in permittivity.

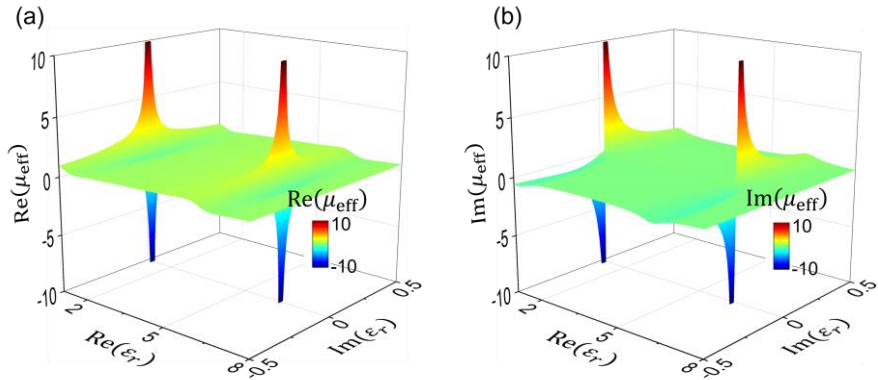

**Figure S3.** (a) Real and (b) imaginary parts of effective relative permeability $\mu_{\text{eff}}$ of a doped ENZ material as functions of the nonmagnetic dopant's relative permittivity $\varepsilon$. The relevant parameters are $A_{\text{total}} = \lambda_0^2$ and $r = 0.35\lambda_0$.

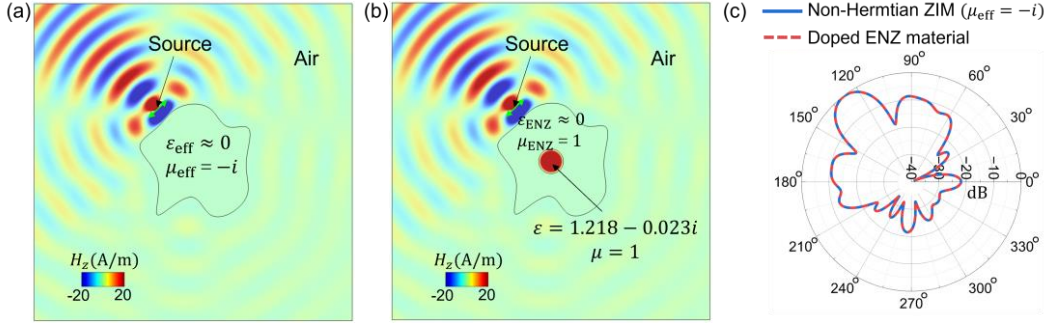

**Figure S4.** [(a) and (b)] Simulated magnetic-field distributions for (a) an non-Hermitian ZIM with effective parameters $\varepsilon_{\text{eff}} \approx 0$ and $\mu_{\text{eff}} = -i$, and (b) a ENZ material ($\varepsilon_{\text{ENZ}} \approx 0$, $\mu_{\text{ENZ}} = 1$) doped with a dielectric dopant with gain ($\varepsilon = 1.218 - 0.023i$, $\mu = 1$) under the illumination of an in-plane electric dipole source. The relevant parameters are $A_{\text{total}} = 7.076\lambda_0^2$ and $r = 0.35\lambda_0$. (c) Far-field radiation patterns for the models in (a) and (b).

Figure S4 provides a numerical example case for verification. First, we consider a desired irregular non-Hermitian ZIM with effective parameters $\varepsilon_{\text{eff}} \approx 0$ and $\mu_{\text{eff}} = -i$ [Fig. S4(a)]. Based on Eq. (S20), we have successfully designed a doped ENZ material exhibiting these effective parameters. Specifically, a dielectric dopant with gain ($\varepsilon = 1.218 - 0.023i$, $\mu = 1$) is doped inside an ENZ material ($\varepsilon_{\text{ENZ}} \approx 0$, $\mu_{\text{ENZ}} = 1$) [Fig. S4(b)]. The simulated magnetic-field distributions for both models under the illumination of an in-plane electric dipole source show excellent agreement. Moreover, the far-field radiation patterns in Fig. S4(c) for both models also perfectly match, confirming the effectiveness of the photonic doping approach.

It is noteworthy that a hybrid doping strategy—combining a lossless dielectric dopant with a lossy/gain dopant—enables independent control over the real and imaginary parts of the effective permeability $\mu_{\text{eff}}$. The lossless dopant can be designed to nullify the real part, while the lossy/gain dopant tailors the imaginary part, yielding purely effective imaginary permeability [2].

In practice, the ENZ materials have been realized in various frequencies from microwaves to terahertz, infrared, and visible spectra [1, 3, 4]. Non-Hermitian ZIMs exhibiting effective positive imaginary permeability have already been experimentally realized using dielectric dopants with loss [5]. Regarding the experimental realization of gain dopants, in the optical and infrared regimes, they can be achieved by doping dielectric materials with fluorescent dyes, quantum wells, or quantum dots [6]. At microwave frequencies, they can be achieved through negative-resistance-integrated metamaterials [7].

**Table 2.** The relevant parameters for the photonic doping models in Fig 2.

|  | Parameters | Lasing EP | Reflecting EP | Absorbing EP |
|---|---|---|---|---|
| **ZIM in** | $r$ | $0.31\lambda_0$ | $0.31\lambda_0$ | $0.31\lambda_0$ |
| **channel 1** | $\varepsilon$ | $1.695 + 0.03i$ | $1.7$ | $1.695 - 0.03i$ |
| **ZIM in** | $r$ | $0.31\lambda_0$ | $0.31\lambda_0$ | $0.31\lambda_0$ |
| **channel 2** | $\varepsilon$ | $1.695 + 0.03i$ | $1.7$ | $1.695 - 0.03i$ |
| **ZIM in** | $r$ | $0.31\lambda_0$ | $0.31\lambda_0$ | $0.31\lambda_0$ |
| **channel 3** | $\varepsilon$ | $1.695 + 0.03i$ | $1.7$ | $1.695 - 0.03i$ |

Based on Eq. (S20), we have designed the photonic doping models for realizing third-order lasing, reflecting, and absorbing EPs, as presented in Figs. 2 and 3 in the main text. The relevant parameters are summarized in Tables 2 and 3. We note that we can directly use an ENZ material for realizing the ZIM CPU with $\mu_C \approx 0$ in Fig. 2, with no need for doping, due to its negligibly small area.

**Table 3.** The relevant parameters for the photonic doping models in Fig 3.

|  | Parameters | Lasing EP | Reflecting EP | Absorbing EP |
|---|---|---|---|---|
| **ZIM in channel 1** | $r$ | $0.373\lambda_0$ | $0.3\lambda_0$ | $0.359\lambda_0$ |
|  | $\varepsilon$ | $1.312 - 0.0231i$ | $1.45$ | $1.353 + 0.095i$ |
| **ZIM in channel 2** | $r$ | $0.373\lambda_0$ | $0.3\lambda_0$ | $0.357\lambda_0$ |
|  | $\varepsilon$ | $1.2692 - 0.0991i$ | $1.862 + 0.04i$ | $1.404 + 0.019i$ |
| **ZIM in channel 3** | $r$ | $0.373\lambda_0$ | $0.35\lambda_0$ | $0.352\lambda_0$ |
|  | $\varepsilon$ | $1.2961 - 0.06645i$ | $1.862 - 0.04i$ | $1.422 + 0.061i$ |
| **ZIM CPU** | $r$ | $0.219\lambda_0$ | $0.219\lambda_0$ | $0.219\lambda_0$ |
|  | $\varepsilon$ | $4$ | $4$ | $4$ |

## 7. Wave phenomena at third-order lasing, reflecting, and absorbing EPs

To clearly show the wave behaviors at the third-order lasing, reflecting, and absorbing EPs, in Fig. S5, we present the normalized magnetic-field amplitudes along the white dashed lines marked within the channels. These models are adopted from Fig. 2 in the main text. At the third-order lasing EP [Fig. S5(a)], the magnetic-field amplitude is significantly larger compared to the incident field, indicating the presence of strong outgoing waves. While at the third-order reflecting EP [Fig. S5(b)], we observe standing waves, confirming that all incident waves are perfectly reflected. At the absorbing EP [Fig. S5(c)], the normalized magnetic-field amplitude is unity, which indicates complete absorption of the incident waves.

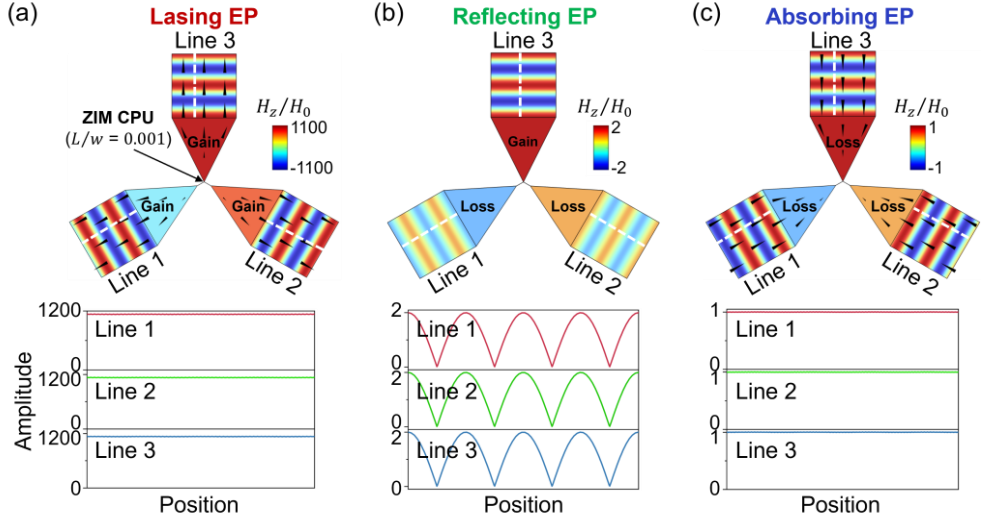

**Figure S5.** The normalized magnetic fields (upper) and magnetic-field amplitudes along the white dashed lines (lower) for the third-order (a) lasing EP, (b) reflecting EP, and (c) absorbing EP. These models are adopted from Fig. 2 in the main text.

## 8.  Model of ultrasensitive gas sensor via non-Hermitian ZIMs

For the gas sensor studied in Fig. 4 in the main text, the central ZIM CPU is implemented using an ENZ material doped with a core-shell dopant, where a dielectric shell surrounds a vacuum core. The vacuum core allows gases to pass through, inducing a refractive index perturbation $\delta n$. Figure S6 provides a top-down view of the gas sensor model. Here, $\varepsilon_c$ (or $\varepsilon_s$) and $r_c$ (or $r_s$) denote the relative permittivity and radius of the core (or shell), respectively. It is important to note that Eq. (S20) does not apply to this core-shell dopant. Therefore, we proceed to derive the effective permeability for the ENZ material doped with the core-shell dopant.

The magnetic field in the ENZ material is uniform which can be expressed as

$$\mathbf{H}_{\text{ENZ}} = H_0 \mathbf{e_z}, \tag{S21}$$

where $H_0$ is the magnetic-field amplitude. In this situation, the magnetic fields within the core and shell can be expressed as follows,

$$\mathbf{H}_c = a_c J_0(k_c r)\mathbf{e_z}, \tag{S22}$$

$$\text{and } \mathbf{H}_s = [b_s J_0(k_s r) + c_s N_0(k_s r)]\mathbf{e_z}, \tag{S23}$$

where $a_c$, $b_s$, and $c_s$ are the unknown coefficients to be determined; $J_m$ and $N_m$ are the $m$th order Bessel and Neumann function, respectively; $k_c$ and $k_s$  represent the wave numbers in the core and shell, respectively. Then, the electric fields within the core and shell can be obtained using Eq. (S2),

$$\mathbf{E}_c = iZ_0 \sqrt{\frac{1}{\varepsilon_c}} a_c J_1(k_c r) \mathbf{e}_{\boldsymbol{\theta}}, \tag{S24}$$

and  $\mathbf{E}_s = iZ_0 \sqrt{\frac{1}{\varepsilon_s}} [b_s J_1(k_s r) + c_s N_1(k_s r)] \mathbf{e}_{\boldsymbol{\theta}}. \tag{S25}$

Considering the boundary continuity condition at the dopant core-shell interface, we have

$$a_c J_0(k_c r_c) = b_s J_0(k_s r_c) + c_s N_0(k_s r_c), \tag{S26}$$

and  $\sqrt{\frac{\mu_c}{\varepsilon_c}} a_c J_1(k_c r_c) = \sqrt{\frac{\mu_s}{\varepsilon_s}} [b_s J_1(k_s r_c) + c_s N_1(k_s r_c)]. \tag{S27}$

Similarly, at the interface between the dopant and the ENZ material, we obtain

$$b_s J_0(k_s r_s) + c_s N_0(k_s r_s) = H_0. \tag{S28}$$

Combining Eqs. (S26)-(S28) yields the effective relative permeability  $\mu_{\text{eff}}$:

$$\mu_{\text{eff}} = \frac{1}{A_{\text{total}}} \left[ A_{\text{total}} - \pi r_s^2 + \frac{2\pi r_s}{\sqrt{\varepsilon_s} k_0} \frac{D_1 J_1(k_s r_s) + D_2 N_1(k_s r_s)}{D_3} \right], \tag{S29}$$

where

$$D_1 = \sqrt{\frac{1}{\varepsilon_c}} N_0(k_s r_c) J_1(k_c r_c) - \sqrt{\frac{1}{\varepsilon_s}} N_1(k_s r_c) J_0(k_c r_c), \tag{S30a}$$

$$D_2 = \sqrt{\frac{1}{\varepsilon_s}} J_0(k_c r_c) J_1(k_s r_c) - \sqrt{\frac{1}{\varepsilon_c}} J_0(k_s r_c) J_1(k_c r_c), \tag{S30b}$$

$$D_3 = \sqrt{\frac{1}{\varepsilon_c}} J_1(k_c r_c) [N_0(k_s r_c) J_0(k_s r_s) - N_0(k_s r_s) J_0(k_s r_c)] + \sqrt{\frac{1}{\varepsilon_s}} J_0(k_c r_c) [N_0(k_s r_s) J_1(k_s r_c) - J_0(k_s r_s) N_1(k_s r_c)]. \tag{S30c}$$

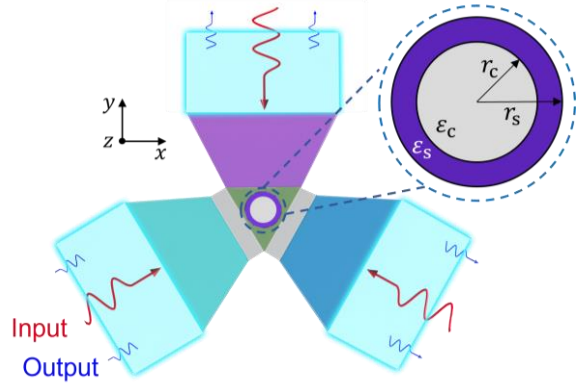

**Figure S6.** Top-down view of gas sensor model. The central ZIM CPU is implemented using an ENZ material doped with a core-shell dopant, where the vacuum core allows gases to pass through.

In the practical implementation, the core of the dopant is set to vacuum with  $\varepsilon_c = 1$ , while the shell is a dielectric material with $\varepsilon_s = 4$. By carefully tuning the radius of the core and shell, an effective permeability $\mu_{\mathrm{eff}} = 0$ can be achieved for the ZIM CPU, enabling the realization of the third-order absorbing EP. Additionally, the non-Hermitian ZIM within each channel can be implemented using a circular dopant according to Eq. (S20). The relevant parameters of the dopants are summarized in Table 4. It is noteworthy that the proposed gas sensor model is highly practical, as all dopants are composed of lossy materials.

**Table 4.** The relevant parameters of dopants for the gas sensor.

| | | |
|---|---|---|
| **ZIM in channel 1** | $r$ | $0.359\lambda_0$ |
| | $\varepsilon$ | $1.353 + 0.095i$ |
| **ZIM in channel 2** | $r$ | $0.357\lambda_0$ |
| | $\varepsilon$ | $1.404 + 0.019i$ |
| **ZIM in channel 3** | $r$ | $0.352\lambda_0$ |
| | $\varepsilon$ | $1.422 + 0.061i$ |
| **ZIM CPU** | $r_c$ | $0.136626\lambda_0$ |
| | $\varepsilon_c$ | $1$ |
| | $r_s$ | $0.25\lambda_0$ |
| | $\varepsilon_s$ | $4$ |

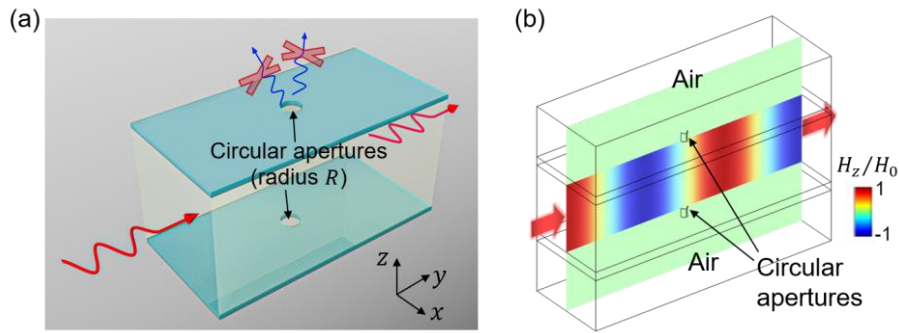

**Figure S7.** (a) Schematic of a parallel-plate waveguide with circular apertures (radius $R$) on its upper and lower plates. (b) Simulated distribution of normalized magnetic field when a transverse electromagnetic mode is excited at the left port and propagates towards the right port.

Finally, we note that while the core where the gas passes through must be open to enable gas flow, radiation losses can be effectively suppressed when the aperture dimensions are deeply subwavelength, as detailed below.

Here, we examine radiation losses in a parallel-plate waveguide with circular apertures (radius $R$) on its upper and lower plates, as depicted in Fig. S7(a). The upper and lower plates are made of perfect magnetic conductors, similar to those in the proposed gas sensor. The waveguide mode within the parallel-plate waveguide has the potential to excite the modes inside the circular apertures, which may lead to wave leakage. Based on circular waveguide theory [8], the cutoff wavelength of the fundamental mode is $2\pi R/1.841$. For the free-space wavelength $\lambda_0$ is significantly larger than the cutoff wavelength, i.e., $\lambda_0 \gg 2\pi R/1.841$, the radiation losses become negligible. This implies that by employing deep-subwavelength apertures that satisfy $R/\lambda_0 \ll 1.841/2\pi \approx 0.293$, we can effectively suppress radiation losses from these apertures while still allowing gases to pass through.

Figure S7(b) presents a numerical example for verification. A transverse electromagnetic mode, with magnetic field along the $z$ direction, is excited at the left port of the parallel-plate waveguide and propagates towards the right port. Two identical apertures, each with a radius $R = 0.02\lambda_0$, are opened on the upper and lower plates. The simulated distribution of normalized magnetic field $H_z/H_0$ ($H_0$ is the magnetic-field amplitude of incidence) is presented in Fig. S7(b), showing almost no wave leakage from the two apertures. Moreover, the waveguide mode remains undistorted. These results confirm that deep-subwavelength apertures can effectively suppress radiation losses.

**References**

[1] I. Liberal, A. M. Mahmoud, Y. Li, B. Edwards, and N. Engheta, Photonic doping of epsilon-near-zero media, Science **355**, 1058-1062 (2017).

[2] W. Ji, D. Wang, S. Li, Y. Shang, W. Xiong, L. Zhang, and J. Luo, Photonic-doped epsilon-near-zero media for coherent perfect absorption, Appl. Phys. A **125**, 129 (2019).

[3] I. Liberal, and N. Engheta, Near-zero refractive index photonics, Nat. Photonics **11**, 149-158 (2017).

[4] X. Niu, X. Hu, S. Chu, and Q. Gong, Epsilon-near-zero photonics: A new platform for integrated devices, Adv. Opt. Mater. **6**, 1701292 (2018).

[5] J. Luo, B. Liu, Z. H. Hang, and Y. Lai, Coherent perfect absorption via photonic doping of zero-index media, Laser Photon. Rev. **12**, 1800001 (2018).

[6]  O. Hess, J. B. Pendry, S. A. Maier, R. F. Oulton, J. M. Hamm, and K. L. Tsakmakidis, Active nanoplasmonic metamaterials, Nat. Mater. **11**, 573-584 (2012).

[7]  C. Qian, Y. Yang, Y. Hua, C. Wang, X. Lin, T. Cai, D. Ye, E. Li, I. Kaminer, and H. Chen, Breaking the fundamental scattering limit with gain metasurfaces, Nat. Commun. **13**, 4383 (2022).

[8]  D. M. Pozar, *Microwave Engineering* (John Wiley & Sons, Inc., 2012), 4 ed.